\documentclass[usenatbib]{mn2e}
\usepackage{psfig}   
\def\dmu{\mbox{pc cm$^{-3}$}}

\title[Radiation properties of extreme nulling pulsar J1502$-$5653]
{Radiation properties of extreme nulling pulsar J1502$-$5653}

\author[J. Li et al.]  {J. Li$^{1,2}$,
  A. Esamdin$^{1}$\thanks{E-mail: aliyi@xao.ac.cn},
  R. N. Manchester$^{3}$, M. F. Qian$^{1,2}$ H. B. Niu$^{1,2}$ \\
  $^{1}$Xinjiang Astronomical Observatory, Chinese Academy of Sciences,
  150, Science 1-street, Urumqi, Xinjiang, 830011, China \\
  $^{2}$Graduate University of Chinese Academy of Sciences, 19A Yuquan Road,
  Beijing, 100049, China \\
  $^{3}$CSIRO Astronomy and Space Science, Australia Telescope
  National Facility, PO Box 76, Epping NSW 1710, Australia \\
}
\date{\today}

\begin{document}
\label{firstpage}
\maketitle

\begin{abstract}
  We report on radiation properties of extreme nulling pulsar
  J1502$-$5653, by analyzing the data acquired from the Parkes 64-m
  telescope at 1374 MHz. The radio emission from this pulsar exhibits
  sequences of several tens to several hundreds consecutive burst
  pulses, separated by null pulses, and the appearance of the emission
  seems quasi-periodic. The null fraction from the data is estimated
  to be 93.6\%. No emission is detected in the integrated profile of
  all null pulses. Systematic modulations of pulse
  intensity and phase are found at the beginning of burst-pulse
  sequences just after null. The intensity usually rises to a maximum
  for the first few pulses, then declines exponentially
  afterwards, and becomes stable after few tens of pulse periods. The
  peak phase appears at later longitudes for the first pulse, then
  drifts to earlier longitudes rapidly, and then systematic drifting
  gradually vanishes while the intensity becomes stable. In this
  pulsar, the intensity variation and phase modulation of pulses are
  correlated in a short duration after the emission starts following a
  null. Observed properties of the pulsar are compared with other
  nulling pulsars published previously, and the possible explanation
  for phase modulation is discussed.
\end{abstract}


\section{Introduction} \label{sec:intro}

Pulsar nulling, which was first reported by \citet{bac70}, is a
phenomenon in which the pulse emission abruptly turns off for a
certain number of pulse periods, then suddenly returns to
normal. Early studies showed that the ``Nulling Fraction'' (NF),
i.e. the fraction of time that a pulsar is in null state, of most
nulling pulsars is less than 10\% \citep{big92a,viv95}.
\citet{wmj07} studied a sample of 23 nulling pulsars,
including some extreme nulling pulsars with NF up to 95\%.

Investigating the emission behaviors of nulling pulsars is important
to understand the pulsar emission mechanism. Different patterns of
transition between null and burst state have been noted by several
authors. For PSR B1749$-$28 \citep{rit76}, B0809$+$74
\citep{la83,vsrr03}, B1944$+$17 \citep{rit76,dchr86} and B0818$-$41
\citep{bgg10}, the onset of burst is abrupt, and the transition from
burst to null state shows a gradual decline of pulse
emission. However, the pulse intensity increases gradually when
emission starts after a null for PSR J0941$-$39 \citep{bb10}, and the
cessation of emission is sudden for PSR B0031$-$07 \citep{viv95} and
B0818$-$13 \citep{la83}. \citet{bgg10} investigated the post- and
pre-null emission behavior of PSR B0818$-$41, and showed that the
first few pulses after the nulls outshine following pulses, whereas
the last few pulses before the nulls are less intense than other
pulses, and they noted that the phenomenon of null may be associated
with some kind of `reset' of the pulsar radio emission engine.  Null
of most pulsars occurs randomly. However, \citet{klo+06} reported the
quasi-periodic nulls of B1931$+$24, furthermore, periodicity in
nulling pulsars has been detected in PSR B1133$+$16 \citep{hr07},
J1819$+$1305 \citep{rw08} and J1738$-$2330 \citep{gjk09}.

PSR J1502$-$5653 was discovered during the Parkes Multibeam Pulsar
Survey \citep{hfs+04}. The rotation period of the pulsar $P$ is
0.535 s, and its first derivative $\dot{P}$ is
1.83$\times{10^{-15}}$ s s$^{-1}$. Correspondingly, it has a
characteristic age of 4.64$\times{10^6}$ years and surface magnetic
field strength of $10^{12}$ gauss \citep{hfs+04}. \citet{wmj07}
investigated J1502$-$5653 at 1518 MHz and showed that this pulsar
has a NF of 93\%, which makes it an extreme nulling pulsar, with
active pulses lasting typically a minute at intervals of 10 to 15
min of null pulses.

In this paper we carry out a detailed investigation of the emission
behavior of PSR J1502$-$5653. Data analysis and results are presented
in Section 2. The implications of the results are discussed in Section
3. Finally in Section 4, we summarize this work.

\section{Data analysis and results}\label{sec_res}

The data were obtained on September 12, 2002 using the Parkes 64-m
telescope, at a central frequency of 1374 MHz. The data last for 6
hours, and contain 40308 pulse periods. The filterbank
system has a total bandwidth of 288 MHz with $96\times3$MHz channels
of polarization-summed data (for each beam) which are sampled every
1 ms. Details of the observing system are described by
\citet{mlc+01}. The single-pulse time sequence is obtained by
de-dispersing the data at a dispersion measure ($DM$) of 194.0 \dmu.
Pulse intensities were computed by summing samples within an on-pulse
window of width 20 ms and subtracting the baseline level determined
in an off-pulse window of width 200 ms.

\subsection{Time sequence and blocks of successive pulses}

\begin{figure}
\centerline{\psfig{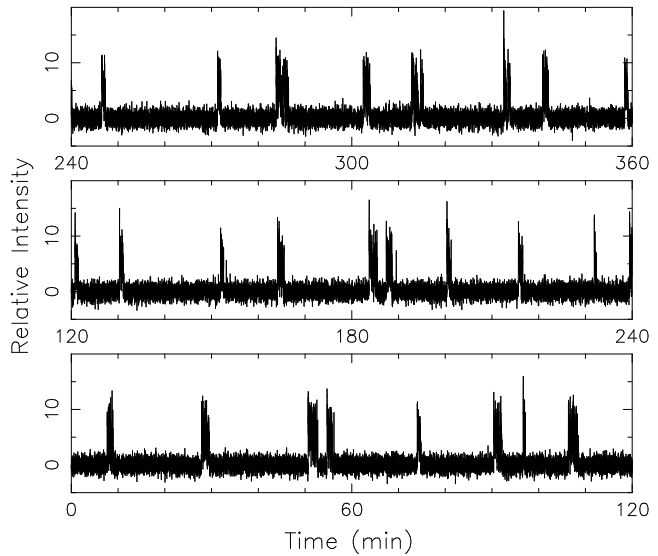}}
\caption{Six hours of time sequence of PSR J1502$-$5653. The time
sequence is equally divided into three panels, each presents two
hours of data.} \label{fig_tsline}
\end{figure}

As shown in Fig. \ref{fig_tsline}, the time sequence shows many
blocks of consecutive strong pulses. In this paper, we considered
intervals more than ten pulse periods with no detectable emission as
null state, and intervals between null states as burst states. In
this way, a total of 29 blocks of burst state (3451 pulses) are
identified. The duration of these blocks varies from about 32 s (60
pulses) to 2 min (240 pulses), with an average duration of 1 min,
while null state lasts from about 16 s (30 pulses) to 25 min (2800
pulses). The Fourier transformation of the autocorrelation function
of the whole time sequence shows two relatively broad peaks at
periods of 11 min and 18 min, implying the burst appearance of the
pulsar may be quasi-periodic.

\begin{figure}
\centerline{\psfig{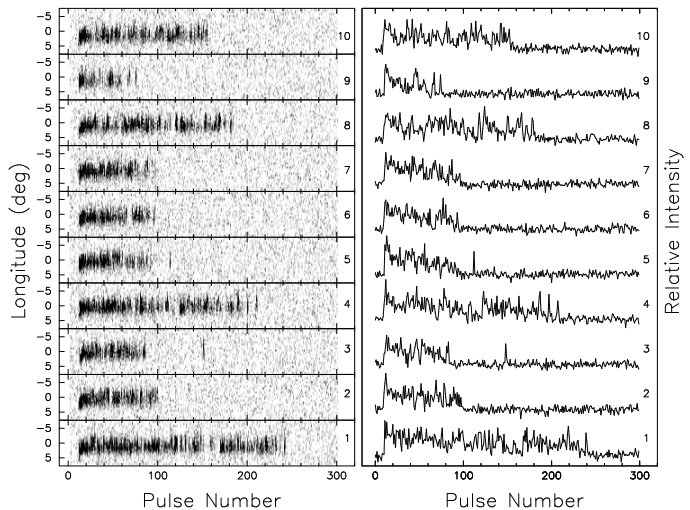}}
\caption{Ten blocks of 300 individual pulse periods containing burst
presented in grayscale diagram (left panel) and intensity diagram
(right panel) from PSR J1502$-$5653. Phase shifting to earlier
longitude and intensity declining can be seen clearly at the early
stage of each burst block.} \label{fig_block10}
\end{figure}

Ten typical blocks of burst are displayed in separate plots of
Fig. \ref{fig_block10} in the form of grayscale diagram (left panel)
and intensity diagram (right panel).  The burst blocks in these plots
begin from the 11th pulse, and preceding 10 nulls are reserved for
comparison. The first ten pulses in each block are quite strong,
and the first few tens of pulses are uninterrupted
by nulls. However, in the middle or late stages of some
burst blocks, the pulse sequence is interrupted by a few short nulls,
usually less than 10 periods. Just following some burst blocks,
one or two sporadic strong pulses are detected occasionally during
the null state. These sporadic single pulses are similar
to the pulses during burst states in intensity, phase and shape.

As can be seen in Fig. \ref{fig_block10}, the pulse intensity shoots
up to a relatively high magnitude for the first few pulses, and then
the pulse intensity drops gradually. After about ten to twenty pulses,
this relatively steady decrease is replaced by a pattern of
random fluctuation in intensity.  As shown in the left panel
of Fig. \ref{fig_block10}, the single pulses drift from later to
earlier longitudes at the beginning of each block of burst,
and then present irregular modulation in pulse phase. These
indicate that the variations of pulse intensity and pulse phase
modulations may be correlated at the early stage of burst; this is
studied further in Section 2.3. All the 29 blocks of burst start
with abrupt rise of the intensity, and at least 23 of them end up
with a gradual decline.

\subsection{Average profiles and pulse energy distributions}

\begin{figure}
\centerline{\psfig{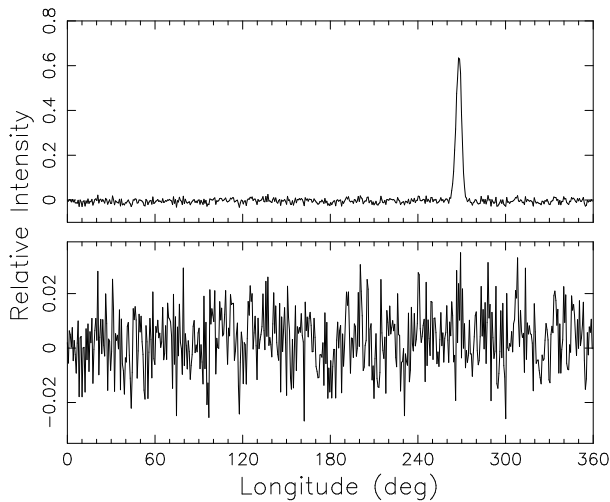}}
\caption{Integrated profiles of PSR J1502$-$5653 for all 40308
pulses
  (top panel) and the 36857 null pulses (bottom panel).}
\label{fig_prof}
\end{figure}

Fig. \ref{fig_prof} shows the integrated profiles for the whole data
span including both burst and null pulses and for just the null
pulses. There is no detectable profile by integrating all pulses in
null state, whereas when the 3451 pulses in burst blocks
added in, the profile is prominent showing that the
burst pulses are actually very strong. The average profile of the
pulsar is narrow, with a 10 per cent width of 9.4$^\circ$ (14 ms) in
longitude.

\begin{figure}
\centerline{\psfig{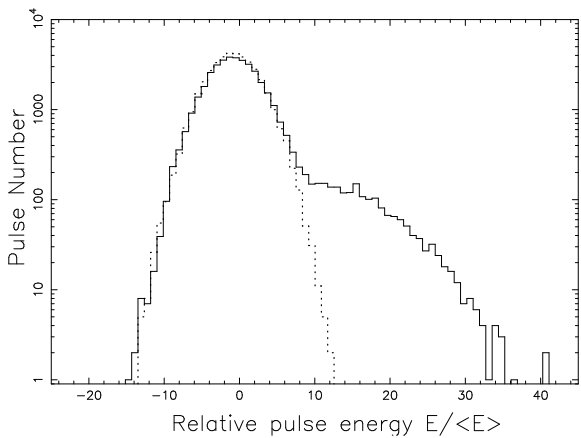}}
\caption{Histogram of on-pulse (solid line) and off-pulse (dotted
line) energies normalized by the mean pulse energy for PSR
J1502$-$5653.} \label{fig_histogram}
\end{figure}

Fig. \ref{fig_histogram} presents histograms of pulse energy
distribution in the pulsar's on-pulse and off-pulse windows, which
are constructed using the method described by \citet{rit76}.
On-pulse and off-pulse energies are determined by integrating within
a on-pulse window and in a off-pulse window of same duration after a
same baseline subtracted, respectively. The histogram formed from
off-pulse energies (dot line) centers around zero, while that from
on-pulse energies (solid line) have a ``long tail" component due to
burst pulses and a big gaussian component due to null pulses.  The
NF of the pulsar is estimated to be 93.6\% through the histograms.
As shown in this figure, the energy of the strongest pulse is 42
times that of mean pulse, suggesting the burst pulses are strong and
highly modulated. This is the first pulse in the burst which
situated at about 332 min in Fig. \ref{fig_tsline}.

\subsection{Single pulse intensity variation and phase modulation}

\begin{figure}
\centerline{\psfig{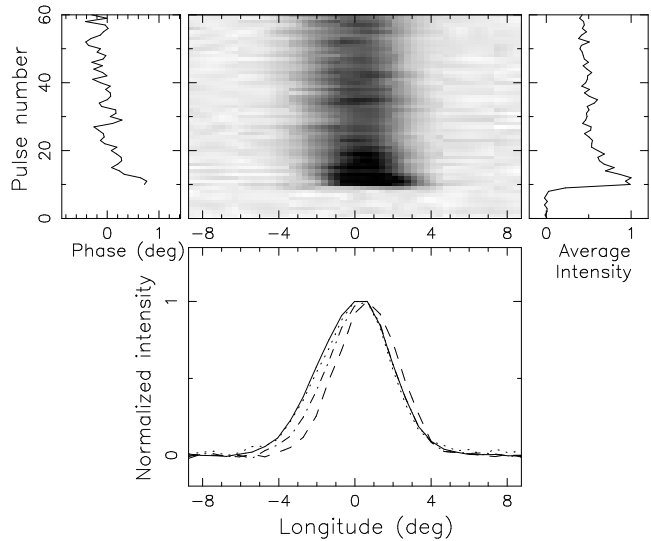}}
\caption{Phase and intensity variations of mean pulse sequence (see
  text). The top-middle panel shows gray-scale plot of the sequence.
  The top-left and top-right panel present phase and intensity
  variation of the mean pulse sequence respectively. The bottom
  panel shows the average profile (solid line) and three profiles,
  which are formed from the first three pulses (dashed line) in
  burst state, the fourth to ninth pulse (dot-dashed line) and
  the tenth to 15th pulse (dotted line), respectively.}
\label{fig_phase}
\end{figure}

To further study the emission behavior of PSR J1502$-$5653 during the
early stage of burst just after null, we construct the mean pulse
sequence by superimposing the first 50 detectable burst pulses of all
blocks (the shortest burst block contains more than 50 burst pulses)
in accordance with the sequence of pulses and the pulse phases,
while ten earlier pulses are also included for comparison.
The distinguishable boundaries from null to burst, the abundance of
burst blocks in the data and no null appears in the first 50
pulses in all burst blocks make this method feasible and effective
in investigating the emission properties of the early stage of burst.
The result is plotted in the top-middle panel of Fig. \ref{fig_phase}.
The intensity fluctuation and phase modulation of the first ten pulses
in the burst state look different from that of the following pulses.

\begin{figure}
\centerline{\psfig{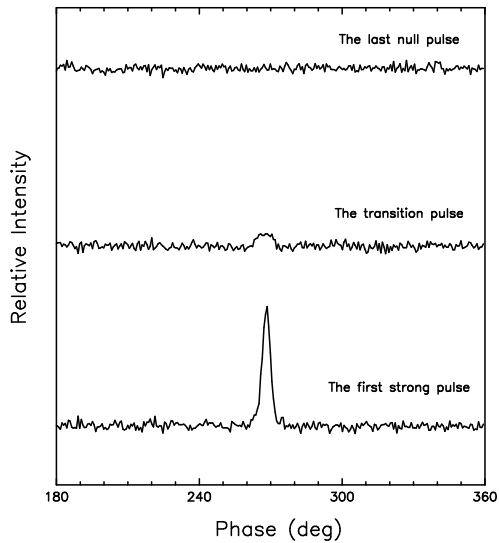}}
\caption{Plots of three consecutive pulses selected from the mean
pulse sequence. From top to bottom, they are the 9th, 10th and 11th
pulses in the mean pulse sequence. The top one is the last null
pulse, the middle is the transition pulse at the start of burst and
the bottom one corresponds to the first pulse after the transition
pulse in the mean pulse sequence.} \label{fig_plot}
\end{figure}

As shown in the top-right panel of Fig. \ref{fig_phase}, the intensity
shoots up at the first mean pulse in the burst state, and remains
strong for about three pulses, then goes down exponentially for following
sequence of about twenty pulses, and becomes stable at the half of maximum
intensity for next tens of pulses.
Using the method described by \citet{bgg10}, we calculate that the
average intensity of the first three mean pulses in the burst state is
1.4, 1.8, 2.2 and 2.4 times that of the following No. 4-9, 10-15, 31-40
and 41-50 mean pulses respectively.

The top-left panel of Fig. \ref{fig_phase} shows that the peak phases
of the first few burst pulses appear at later longitudes than that of the
following pulses. In about thirteen pulse periods, the pulse phase
drifts about 0.8 degree to earlier longitudes, and the phase
of the 13th pulse is equal to the peak phase of average profile,
then the apparent drifting stops and is replaced by irregular phase
modulation. We note that the intensity fluctuation and phase
modulation of the pulsar is correlated in the beginning of burst.

In Fig. \ref{fig_phase} there is some evidence for a weak wide pulse
at the start of burst. This pulse is the 10th pulse in the mean
pulse sequence. We call it ``transition pulse'' in this paper. Fig.
\ref{fig_plot} shows three consecutive pulses selected from the mean
pulse sequence, the last null pulse, the transition pulse, and the
first strong pulse. The transition pulse have the full width at half
maximum (FWHM) of 6.7$^\circ$ and signal-to-noise ratio (S/N) of
4.34. For comparison, the S/N of the first strong burst pulse is
44.1 and the width is 4.04$^\circ$. The transition pulse is very
weak, so it is only detectable in the mean pulse sequence.

The bottom panel in Fig. \ref{fig_phase} displays the average profile
(solid line) of the pulsar and three profiles which are obtained from
the first three pulses following the transition pulse,
the fourth to the ninth pulses and the tenth to 15th pulses, respectively.
The peaks of these three profiles appear at longitudes 0.72$^\circ$,
0.35$^\circ$ and 0.13$^\circ$ respectively, where the peak phase of
the average profile is set as zero. The widths of these three
profiles are 3.86$^\circ$, 4.06$^\circ$ and 4.33$^\circ$, respectively,
while that of the average profile is 4.53$^\circ$.

\begin{figure}
\centerline{\psfig{file=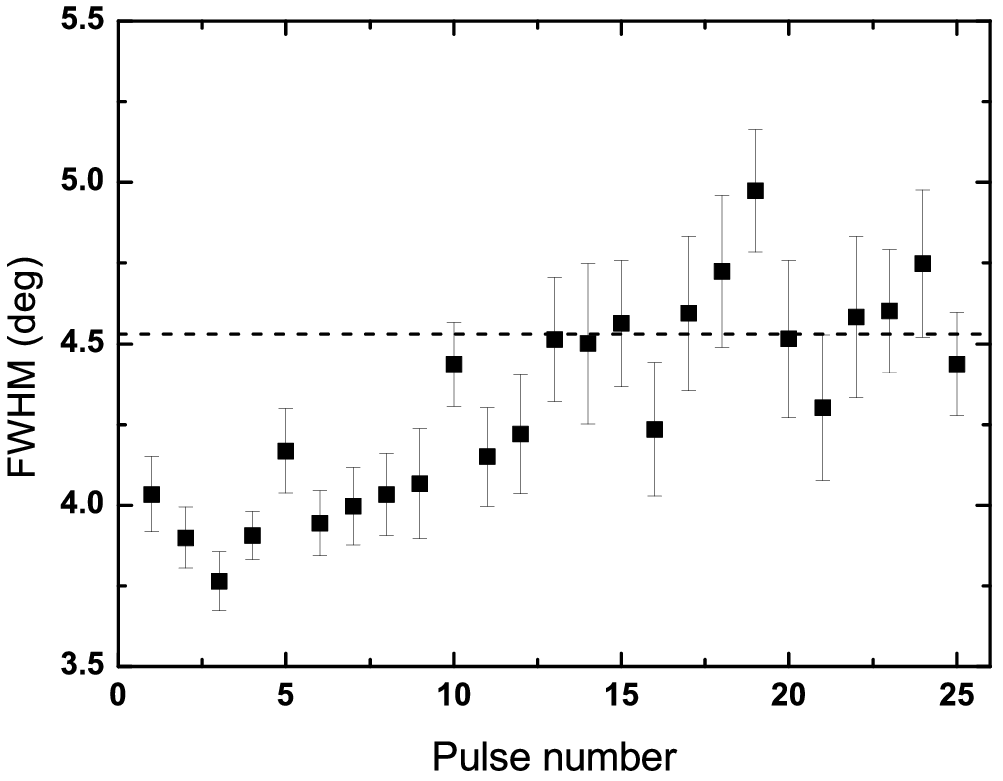,width=65mm,angle=0}} \caption{FWHM
of the 25 mean pulses in burst state of the mean pulse sequence
versus the corresponding pulse number. The dashed line denotes the
pulse width of the average profile. The error of the width is
derived from the uncertainty given by standard guassian fitting
procedure.} \label{fig_width}
\end{figure}

Fig. \ref{fig_width} presents the FWHM of \textbf{25} mean pulses
in burst state, excluding the transition pulse, showing that
the FWHM increases with pulse number at the beginning
of burst. Around the 13th pulse of the burst the width reaches that of
the average profile.
Apart from the wide transition pulse, it is clear that in a short
duration after the null, the radiation window gradually
broadens while the pulses drift from later to earlier longitudes.
The middle and later stage in burst block are often
randomly disrupted by short nulls, and the behaviour is not so
systematic as in the early stages.

\section{DISCUSSION} \label{sec:dis}

The emission of J1502$-$5653 is characterized by abrupt transition
from null to burst with a timescale of less than two pulse periods, a
gradual decrease of pulse intensity in the early stage of a burst
which is accompanied by a shift to earlier longitude in pulse phase,
and a broadening in pulse width. Gradual cessation of the emission in
some bursts before nulls is also noticed. Similar behavior can be seen
in PSR B0818$-$14. \citet{bgg10} reported that, for this pulsar, the
transition from nulls to bursts is abrupt and pulse intensity from
bursts to nulls appears to reduce gradually, and the profile shape of
the first few pulses differs from that of average profile. They also
mentioned that the behavior of subpulse drifting at the beginning of
burst pulses after null is different from the following pulses.
Similarly, in PSR J1502$-$5653, the pattern of phase modulation
of the first few pulses is distinct from that of the later pulses
during burst.

Recently, \citet{bb10} discovered bizarre emission behavior of PSR
J0941$-$39, i.e.  sometimes it only emits sporadic pulses and at other
times it behaves just like a nulling pulsar. From Fig. 5 of
\citet{bb10}, we notice that the post-null pulse phases seem to shift
towards earlier longitudes, and this looks similar to the drifting
behavior of PSR J1502$-$5653. However, unlike PSR J1502$-$5653, the
pulse intensity of PSR J0941$-$39 appears to increase gradually at the
beginning of post-null emission, when it behaves like a nulling
pulsar.

The post-null pulse drifting of PSR J1502$-$5653 may be explained by
the vacuum gap model \citep{rs75}. According to the classical vacuum
gap model, the sub-beams of emission circulate around the magnetic
axis, as a result of ${\bf E}\times{\bf B}$ drift of spark plasma
filaments. At the beginning of each burst, the electric field in the
accelerating gap is relatively high and in consequence the observed
${\bf E}\times{\bf B}$ drift-rate is high. Then the sparking process
starts, which produces not only strong radio emission but also
$e^{+}$ $e^{-}$ pairs. Few pulses periods later, the $e^{+}$ $e^{-}$
pairs accumulate in the accelerating gap and decrease the electric
field to a relatively stable value where sparking and radio
emission go on but the drift-rate reduces to nearly zero.  As the
accumulation proceeds the gap electric field strength keeps on
weakening until sparking process breaks down and the radio emission
ceases. \citet{gmg03,ghm+08} refined the vacuum model by introducing
a thermal ion outflow from the hot polar cap surface.
However, the intensity variations and their correlation with
phase modulation of burst pulses at the beginning of burst blocks
need to be further investigated.

\citet{klo+06} noted that PSR B1931$+$24 turns `on' for 5$-$10 days
and `off' for 25$-$35 days, the switch occurs in a quasi-periodic
fashion, and no obvious emission can be found in the integrated
profile of null state. The difference of the slow-down rates in the
`on' and `off' states of this pulsar indicates a massive change in
magnetospheric currents.  The quasi-periodic transition between `on'
and `off' state and non-detection of integrated energy by folding many
null pulses of PSR J1502$-$5653, suggest that the emission behavior of
this pulsar is somehow similar to that of the intermittent pulsar
B1931$+$24, but with very different `on' and `off' timescales. The
scenario of two slow-down rates of PSR B1931$+$24 may be applicable to
PSR J1502$-$5653, however, measuring two slow-down rates is not
possible for this pulsar, because of the very short durations of `on' and
`off' states. The timescale of the magnetospheric-current changing is
believed to be very short, and it is not yet clear whether and how the
pulses emitted at the very beginning of strong burst state are
influenced by the switching process of the magnetospheric currents.

\section{CONCLUSION} \label{sec:con}

The bursts of pulses of PSR J1502$-$5653 have a typical
duration of 1 min or about 100 pulse periods, and they are separated
by nulls lasting from 30 to 2800 pulse periods. The power spectra of
pulse sequence shows two broad peaks at periods of 11 min and 18 min,
revealing that the appearance of the emission may be quasi-periodical
in this pulsar. The nulling fraction estimated from the data is
93.6\%.  The integrated profile of all null pulses shows no emission.
Interestingly, by integrating over 29 pulse sequences, a weak
and wide pulse is found just before the first detected single pulse
in all burst blocks.

At the beginning of burst after null the intensity usually rises to
the maximum immediately, and keeps a high intensity for few pulses,
then the intensity of next twenty or thirty pulses declines
exponentially, and gradually becomes stable at the half of the maximum
intensity. In most case, the cessation of radiation is gradual. The
peak phase of the first pulse in burst usually appears to be at later
longitudes than that of average profile, then the phase drifts quickly
to earlier longitudes for next several pulses.  The drifting then
tends to slow down in next twenty to thirty pulse periods, and then is
replaced by irregular phase modulation. As the peak phase drifts to
earlier longitudes, the pulse intensity declines, meanwhile, the
radiation window broadens gradually till reaching the width of average
profile. A good correlation can be seen between intensity variation
and phase modulation in the early stage of post-null emission.

The phase modulation may be explained by electric field shielding
caused by $e^{+}$ $e^{-}$ pairs produced by sparking process. The
emission behaviors of individual pulses during the transition between
null and burst state may provide very important clue to understand the
underlying switching mechanism of nulling pulsars.

\subsection*{ACKNOWLEDGMENTS}

This work was funded by the National Natural Science Foundation of
China (NSFC) under No.10973026. We thank members of the Pulsar Group
at XAO for helpful discussions. The Parkes radio telescope is part
of the Australia Telescope which is funded by the Commonwealth
Government for operation as a National Facility managed by the
Commonwealth Scientific and Industrial Research Organization.

\bibliographystyle{mn2e}
\bibliography{journals,modrefs,psrrefs,crossrefs}

\begin{thebibliography}{}

\bibitem[\protect\citeauthoryear{Backer}{Backer}{1970}]{bac70}
Backer D.~C.,  1970, Nature, 228, 42

\bibitem[\protect\citeauthoryear{{Bhattacharyya}, {Gupta} \&
  {Gil}}{{Bhattacharyya} et~al.}{2010}]{bgg10}
{Bhattacharyya} B.,  {Gupta} Y.,    {Gil} J.,  2010, \mnras, 408, 407

\bibitem[\protect\citeauthoryear{Biggs}{Biggs}{1992}]{big92a}
Biggs J.~D.,  1992, ApJ, 394, 574

\bibitem[\protect\citeauthoryear{{Burke-Spolaor} \& {Bailes}}{{Burke-Spolaor}
  \& {Bailes}}{2010}]{bb10}
{Burke-Spolaor} S.,  {Bailes} M.,  2010, MNRAS, 402, 855

\bibitem[\protect\citeauthoryear{Deich, Cordes, Hankins \& Rankin}{Deich
  et~al.}{1986}]{dchr86}
Deich W. T.~S.,  Cordes J.~M.,  Hankins T.~H.,    Rankin J.~M.,  1986, ApJ,
  300, 540

\bibitem[\protect\citeauthoryear{{Gajjar}, {Joshi} \& {Kramer}}{{Gajjar}
  et~al.}{2009}]{gjk09}
{Gajjar} V.,  {Joshi} B.~C.,    {Kramer} M.,  2009, in {D.~J.~Saikia,
  D.~A.~Green, Y.~Gupta, \& T.~Venturi} ed., The Low-Frequency Radio Universe
  Vol.~407 of Astronomical Society of the Pacific Conference Series, {Peculiar
  Nulling in PSR J1738-2330}.
p.~304

\bibitem[\protect\citeauthoryear{{Gil}, {Haberl}, {Melikidze}, {Geppert},
  {Zhang} \& {Melikidze} Jr.}{{Gil} et~al.}{2008}]{ghm+08}
{Gil} J.,  {Haberl} F.,  {Melikidze} G.,  {Geppert} U.,  {Zhang} B.,
  {Melikidze} Jr. G.,  2008, \apj, 686, 497

\bibitem[\protect\citeauthoryear{{Gil}, {Melikidze} \& {Geppert}}{{Gil}
  et~al.}{2003}]{gmg03}
{Gil} J.,  {Melikidze} G.~I.,    {Geppert} U.,  2003, \aap, 407, 315

\bibitem[\protect\citeauthoryear{{Herfindal} \& {Rankin}}{{Herfindal} \&
  {Rankin}}{2007}]{hr07}
{Herfindal} J.~L.,  {Rankin} J.~M.,  2007, \mnras, 380, 430

\bibitem[\protect\citeauthoryear{{Hobbs}, {Faulkner}, {Stairs}, {Camilo},
  {Manchester}, {Lyne}, {Kramer}, {D'Amico}, {Kaspi}, {Possenti}, {McLaughlin},
  {Lorimer}, {Burgay}, {Joshi} \& {Crawford}}{{Hobbs} et~al.}{2004}]{hfs+04}
{Hobbs} G.,  {Faulkner} A.,  {Stairs} I.~H.,  {Camilo} F.,  {Manchester} R.~N.,
   {Lyne} A.~G.,  {Kramer} M.,  {D'Amico} N.,  {Kaspi} V.~M.,  {Possenti} A.,
  {McLaughlin} M.~A.,  {Lorimer} D.~R.,  {Burgay} M.,  {Joshi} B.~C.,
  {Crawford} F.,  2004, MNRAS, 352, 1439

\bibitem[\protect\citeauthoryear{{Kramer}, {Lyne}, {O'Brien}, {Jordan} \&
  {Lorimer}}{{Kramer} et~al.}{2006}]{klo+06}
{Kramer} M.,  {Lyne} A.~G.,  {O'Brien} J.~T.,  {Jordan} C.~A.,    {Lorimer}
  D.~R.,  2006, Science, 312, 549

\bibitem[\protect\citeauthoryear{Lyne \& Ashworth}{Lyne \&
  Ashworth}{1983}]{la83}
Lyne A.~G.,  Ashworth M.,  1983, MNRAS, 204, 519

\bibitem[\protect\citeauthoryear{Manchester, Lyne, Camilo, Bell, Kaspi,
  D'Amico, McKay, Crawford, Stairs, Possenti, Morris \& Sheppard}{Manchester
  et~al.}{2001}]{mlc+01}
Manchester R.~N.,  Lyne A.~G.,  Camilo F.,  Bell J.~F.,  Kaspi V.~M.,  D'Amico
  N.,  McKay N. P.~F.,  Crawford F.,  Stairs I.~H.,  Possenti A.,  Morris
  D.~J.,    Sheppard D.~C.,  2001, MNRAS, 328, 17

\bibitem[\protect\citeauthoryear{{Rankin} \& {Wright}}{{Rankin} \&
  {Wright}}{2008}]{rw08}
{Rankin} J.~M.,  {Wright} G.~A.~E.,  2008, \mnras, 385, 1923

\bibitem[\protect\citeauthoryear{Ritchings}{Ritchings}{1976}]{rit76}
Ritchings R.~T.,  1976, MNRAS, 176, 249

\bibitem[\protect\citeauthoryear{Ruderman \& Sutherland}{Ruderman \&
  Sutherland}{1975}]{rs75}
Ruderman M.~A.,  Sutherland P.~G.,  1975, ApJ, 196, 51

\bibitem[\protect\citeauthoryear{{van Leeuwen}, {Stappers}, {Ramachandran} \&
  {Rankin}}{{van Leeuwen} et~al.}{2003}]{vsrr03}
{van Leeuwen} A.~G.~J.,  {Stappers} B.~W.,  {Ramachandran} R.,    {Rankin}
  J.~M.,  2003, A\&A, 399, 223

\bibitem[\protect\citeauthoryear{{Vivekanand}}{{Vivekanand}}{1995}]{viv95}
{Vivekanand} M.,  1995, mnras, 274, 785

\bibitem[\protect\citeauthoryear{{Wang}, {Manchester} \& {Johnston}}{{Wang}
  et~al.}{2007}]{wmj07}
{Wang} N.,  {Manchester} R.~N.,    {Johnston} S.,  2007, MNRAS, 377, 1383

\end{thebibliography}
\label{lastpage}
\end{document}